\documentstyle[sprocl,epsf]{article}

\input{psfig}

\def\Journal#1#2#3#4{{#1} {\bf #2}, #3 (#4)}

\def\NPB{{\em Nucl. Phys.} B}
\def\PLB{{\em Phys. Lett.}  B}
\def\PRL{\em Phys. Rev. Lett.}
\def\PRD{{\em Phys. Rev.} D}

\def\be{\begin{equation}}
\def\ee{\end{equation}}
\def\bea{\begin{eqnarray}}
\def\eea{\end{eqnarray}}
\catcode`\@=11
\def\slash{\mathpalette\make@slash}
\def\make@slash#1#2{\setbox\z@\hbox{$#1#2$}%
  \hbox to 0pt{\hss$#1/$\hss\kern-\wd0}\box0}
\catcode`\@=12 

\begin{document}

\begin{flushright}
{\bf UCSD/PTH 98-33}\\
{\bf hep-ph/9809431}\\
{\bf September 1998}\\
\end{flushright}
\vspace{1cm}

\title{Applications of Perturbative NRQCD\footnotemark}

\footnotetext{${}^a\!$ Talk given at the 3rd Workshop on Continuous
Advances in QCD, Minneapolis, 16--19 April 1998. This writeup also
covers the topics presented in the talk ``New Results on Heavy $Q\bar
Q$ Pair Production Close to Threshold'' given at the 29th International
Conference on High Energy Physics, Vancouver, Canada, 23--29 July
1998. 
}

\author{A. H. Hoang}

\address{
Department of Physics, University of California, San Diego,\\
 La Jolla, CA 92093--0319, USA\\ 
E-mail: hoang@einstein.ucsd.edu}


\maketitle\abstracts{
We review the theoretical ideas and tools required to arrive at a
next-to-next-to-leading order (NNLO) description of the heavy
quark-antiquark production cross section in $e^+e^-$ annihilation for
the case that the center of mass kinetic energy of the quarks is
larger than $\Lambda_{\tiny\rm QCD}$. In
this case the NNLO cross section can be calculated with purely
perturbative methods. We present details of the calculation and
discuss two applications, the determination of the bottom quark mass
from $\Upsilon$ sum rules and the top-antitop production cross section
close to threshold in $e^+e^-$ annihilation. 
}
\section{Introduction}
\label{sectionintroduction}
Within the last year significant progress has been achieved in the
perturbative description of heavy quark-antiquark ($Q\bar Q$) pairs in
the kinematic regime close to threshold using the concepts of effective
field theories. In this talk we report on some of these developments
from the point of view of practical applications. There have also been
many interesting new results concerning the systematics of the
effective field theory description of $Q\bar Q$ pairs at threshold
which would deserve to be presented in detail but
can only be mentioned peripherally due to lack of space.

To be definite, we consider the total cross section $\sigma^{thr}
\equiv \sigma(e^+ e^-\to
Q\bar Q + \mbox{anything})$ for the c.m. energies $\sqrt{s}\approx 2
M_Q$, $M_Q$ being the heavy quark mass. In this kinematic regime the
$Q\bar Q$ dynamics is characterized by the fact that the heavy quarks
have small c.m. velocities. Considering the strong dependence of the
velocity ($v=\sqrt{1-4M_Q^2/s}$) in the threshold regime on $M_Q$ it
is obvious that, at least in principle, accurate calculations of the
cross section may lead to precise determinations of the quark
mass. The situation, however, is not simple due to
the fact that the $Q\bar Q$ system in the nonrelativistic regime is
governed by (at least) three scales. This makes this system
particularly complicated. The three scales are $M_Q$, the quark mass,
$M_Q v$, the relative momentum, and $M_Q v^2$, the kinetic energy in
the c.m. frame. This has three important consequences which are
symptomatic for all nonrelativistic $Q\bar Q$ systems:

1. Because $v\ll 1$, the three scales are widely separated ($M_Q\gg
M_Q v\gg M_Q v^2$). The theoretical tools required to describe the
$Q\bar Q$ pair strongly depend on whether the scales each are smaller
or larger then $\Lambda_{\tiny\rm QCD}$. The most transparent situation
arises if $M_Q v^2 > \Lambda_{\tiny\rm QCD}$ because in
this case the theoretical tools resemble most closely those of QED
bound state calculations. In fact, for the two applications mentioned
in this talk it is ensured that this condition is satisfied. 

2. Because ratios of the three scales arise, the description of the
nonrelativistic $Q\bar Q$ pair involves a double expansion in
$\alpha_s$ and $v$. This means that the standard multi-loop expansion
in $\alpha_s$ breaks down. The most prominent indication of this fact
is the so called {\it Coulomb singularity}, originating from the ratio
$M_Q/M_Q v$, which corresponds to a
singular $(\alpha_s/v)^n$-behavior in the n-loop correction to the
amplitude $\gamma\to Q\bar Q$ for $v\to 0$. The latter singularity is
caused by the $00$-component of the gluon propagator and, in Coulomb
gauge, is directly related to the exchange of longitudinally polarized
gluons. This singularity has to be treated by resumming diagrams
involving longitudinal gluon exchange to all orders in $\alpha_s$. The
Coulomb singularity exists in the nonrelativistic limit, but there are
also power-like and logarithmic divergences in $v$ which are
suppressed by powers of the velocity. At this point we would like to
specify more clearly what ``resummation'' means in this
context. Strictly speaking it would mean a resummation of the
perturbation series in $\alpha_s$ to all orders, where the
respective coefficients are expanded up to a certain power in
$v$. This means that the resummation would be carried out in the
(formal) limit $\alpha_s\ll v\ll 1$. The resulting series would then
(uniquely) define analytic functions which could be continued to the
region of interest $|v|\sim\alpha_s$. Typical structures like $Q\bar
Q$ bound states can only be observed after this continuation. Of
course, this would be a highly cumbersome and inefficient method. It is
therefore mandatory to reformulate the problem in terms of wave
equations. The solutions of these wave equations are equivalent to the
results of the 
resummation method. As a matter of convenience we will also call the
wave equation method ``resummation'' for the rest of this talk.

3. From point 1 we can conclude that the c.m. velocity of the heavy
quarks should satisfy the condition $v>(\Lambda_{\tiny\rm
QCD}/M_Q)^{1/2}$ to ensure that the scale $M_Q v^2$ is
perturbative. In addition, relativistic corrections are not
suppressed by factors of $\pi$ like multi-loop corrections. This
means that relativistic corrections can be quite sizeable. 
The corrections of ${\cal{O}}(v^2)$, called NNLO from now on, can be
estimated to be of order $20-30\%$ for $b\bar b$ and $5\%$ for $t\bar
t$. Thus, the calculation of higher order relativistic corrections is
mandatory in order to achieve sufficient theoretical accuracy and to
test the reliability of the perturbative description itself. Obviously
the perturbative treatment works better if $M_Q$ is large.

In order to arrive at a reliable theoretical description of
$\sigma^{\tiny thr}$ we have to go through two steps: first, we have
to address the question how to organize the calculation systematically
keeping in mind points 1-3, and, second, we have to actually carry out
the calculation itself. In Section~\ref{sectionsystematics} we will
briefly address the systematics and in
Section~\ref{sectioncalculation} we will present the calculation of
the photon mediated cross section at
NNLO in the nonrelativistic
expansion. Section~\ref{sectionapplications} is devoted to two
applications, the determination of the bottom quark mass and the
$t\bar t$ cross section at threshold in $e^+e^-$ annihilation.
\section{Systematics and NRQCD}
\label{sectionsystematics}
A very economical approach to systematically deal with the problems
described previously is to take advantage of the separation
of the scales $M_Q$, $M_Q v$ and $M_Q v^2$ using the concepts of
effective field theories. In the following we outline the conceptual
steps to arrive at a NNLO description of a nonrelativistic
$Q\bar Q$ pair for $\Lambda_{\tiny\rm QCD} < M_Q v^2$
without going very far into formal considerations. The basic idea of
the effective field theory approach is to integrate out momenta
above the scales relevant for the nonrelativistic dynamics of the
$Q\bar Q$ pair. Doing this, one always has to keep in mind the
relation of each of the scales to $\Lambda_{\tiny\rm QCD}$. (If $M_Q$
were of order as or even smaller than $\Lambda_{\tiny\rm QCD}$ a
nonrelativistic expansion would be meaningless in the first place.)

Suppose that $M_Q$ is larger than $\Lambda_{\tiny\rm QCD}$. In this
case we can integrate out momenta of order $M_Q$ because they are
not responsible for the nonrelativistic dynamics of the $Q\bar Q$
pair. Starting from QCD we then arrive at an effective field theory in
which the heavy quarks and the gluons interacting with them only carry
momenta below $M_Q$. This forces us to introduce different fields
for heavy quark and antiquark. The resulting theory is called
nonrelativstic QCD (NRQCD)~\cite{Caswell1} and its Lagrangian reads
\begin{eqnarray}
{\cal{L}}_{\mbox{\tiny NRQCD}} & = &
- \frac{1}{2} \,\mbox{Tr} \, G^{\mu\nu} G_{\mu\nu} 
+ \sum_{\tiny\rm light \,\,quarks} \bar q \, i \slash{D} \, q
+\, \psi^\dagger\,\bigg[\,
i D_t 
+ a_1\,\frac{{\mbox{\boldmath $D$}}^2}{2\,M_Q} 
\,\bigg]\,\psi + \ldots 
\nonumber\\[2mm] & &
+ \,\psi^\dagger\,\bigg[\, 
+ a_2\,\frac{{\mbox{\boldmath $D$}}^4}{8\,M_Q^3} +
\frac{a_3\,g}{2\,M_Q}\,{\mbox{\boldmath $\sigma$}}\cdot
    {\mbox{\boldmath $B$}}
+ \, \frac{a_4\,g}{8\,M_Q^2}\,(\,{\mbox{\boldmath $D$}}\cdot 
  {\mbox{\boldmath $E$}}-{\mbox{\boldmath $E$}}\cdot 
  {\mbox{\boldmath $D$}}\,)
 \,\bigg]\,\psi +
\nonumber\\[2mm] & &
+ \,\psi^\dagger\,\bigg[\,
  \frac{a_5\,g}{8\,M_Q^2}\,i\,{\mbox{\boldmath $\sigma$}}\,
  (\,{\mbox{\boldmath $D$}}\times 
  {\mbox{\boldmath $E$}}-{\mbox{\boldmath $E$}}\times 
  {\mbox{\boldmath $D$}}\,)
 \,\bigg]\,\psi 
+\ldots
+ \mbox{$\chi^\dagger\chi$ bilinears}
\,.
\label{NRQCDLagrangian}
\end{eqnarray}
The gluonic and light quark degrees of freedom are described by the
conventional relativistic Lagrangian, whereas the heavy quark and
antiquark are described by the Pauli spinors $\psi$ and $\chi$,
respectively. For convenience all color indices are suppressed. 
Only those terms relevant for the NNLO cross section are displayed,
where we have omitted the straightforward antitop bilinears. The
latter can be obtained through charge conjugation symmetry.
The effects coming from momenta or order $M_Q$ are encoded in the
short-distance coefficients $a_1,\ldots,a_5$. They can be determined
as a perturbative series in $\alpha_s$ at the scale $\mu_{\rm
hard}=M_Q$ through the matching procedure. If $\Lambda_{\tiny\rm QCD}$
were of order of or even larger than $M_Q v$ (which is essentially the
situation for $c\bar c$) this would be all we could do
using perturbation theory. For $\Lambda_{\tiny\rm QCD}$ smaller than
$M_Q v$,
however, one can go further and also integrate out gluonic (and light
quark) momenta of order $M_Q v$. The resulting theory has been
called ``potential NRQCD'' (PNRQCD) in Ref.~\cite{Pineda1} and is
characterized by
the fact that its Lagrangian contains spatially non-local four-fermion
interactions which are nothing else than instantaneous (static) $Q\bar
Q$ potentials. The ``short-distance'' coefficients of the
corresponding operators describe the gluonic (and light quark) effects
from momenta of order $M_Q v$ and can be calculated perturbatively at
the scale $\mu_{\rm soft}=M_Q v$. To NNLO
(i.e.\ including potentials suppressed by at most $\alpha_s^2$,
$\alpha_s/M_Q$ or $1/M_Q^2$ relative to the Coulomb potential) the
relevant $Q\bar Q$ potentials
read ($a_s\equiv\alpha_s(\mu_{\rm soft})$, $C_A=3$, $C_F=4/3$,
$T=1/2$, $\tilde\mu\equiv e^\gamma\,\mu_{\rm soft}$, $r\equiv | \vec r
|$) 
\begin{eqnarray}
V_c(\vec r) & = & -\,\frac{C_F\,a_s}{r}\,
\bigg\{\, 1 +
\Big(\frac{a_s}{4\,\pi}\Big)\,\Big[\,
2\,\beta_0\,\ln(\tilde\mu\,r) + a_1
\,\Big]
\nonumber\\[2mm] & & 
 + \Big(\frac{a_s}{4\pi}\Big)^2\Big[\,
\beta_0^2\Big(4\ln^2(\tilde\mu\,r) 
      + \frac{\pi^2}{3}\Big) 
+ 2\Big(2\beta_0\,a_1 + \beta_1\Big)\ln(\tilde\mu\,r) 
+ a_2
\Big]
\bigg\}
\,,
\label{Coulombpotential}
\\[2mm]
V_{\mbox{\tiny BF}}(\vec r) & = & 
\frac{C_F\,a_s\,\pi}{M_Q^2}\,
\Big[\,
1 + \frac{8}{3}\,\vec S_t \,\vec S_{\bar t}
\,\Big]
\,\delta^{(3)}(\vec r)
+ \frac{C_F\,a_s}{2\,M_Q^2 r}\,\Big[\,
\vec\nabla^2 + \frac{1}{r^2} \vec r\, (\vec r \, \vec\nabla) \vec\nabla
\,\Big]
\nonumber\\[2mm] & &
- \,\frac{3\,C_F\,a_s}{M_Q^2\,r^3}\,
\Big[\,
\frac{1}{3}\,\vec S_t \,\vec S_{\bar t} - 
\frac{1}{r^2}\,\Big(\vec S_t\,\vec r\,\Big)
\,\Big(\vec S_{\bar t}\,\vec r\,\Big)
\,\Big]
+ \frac{3\,C_F\,a_s}{2\,M_Q^2\,r^3}\,\vec L\,(\vec S_t+\vec S_{\bar t})
\label{BFpotential}
\,,\\[2mm]
V_{\mbox{\tiny NA}}(\vec r) & = &
-\,\frac{C_A\,C_F\,a_s^2}{2\,M_Q\,r^2} 
\,,
\label{NApotential}
\end{eqnarray}
where $\vec S_t$ and $\vec S_{\bar t}$ are the top and antitop spin
operators, $\vec L$ is the angular momentum operator and $\beta_{0,1}$
are the one- and two-loop beta-functions. The constants 
$a_{1,2}$ have been calculated in Refs.~\cite{Fischler1,Peter1}. $V_c$
is the Coulomb (static) potential and
$V_{\mbox{\tiny BF}}$ the Breit-Fermi potential known from
positronium. $V_{\mbox{\tiny NA}}$ is a purely non-Abelian potential
generated through non-analytic terms in the one-loop vertex
corrections to the Coulomb potential involving the triple gluon
vertex. 
The remaining dynamical gluon (light quark) fields can only carry
momenta of order $M_Q v^2$ and describe radiation and retardation
effects. If $\Lambda_{\tiny\rm QCD} < M_Q v^2$ one can show 
that those retardation effects are of NNNLO in
$\sigma^{thr}$ using arguments known from QED and taking into account
how the gluon self coupling scales with $v$ for gluonic momenta of
order $M_Q v^2$. [This only works because the $Q\bar Q$ pair is
produced in a color singlet state!] This means that retardation
effects (and the scale $M_Q v^2$) can be ignored
at NNLO and that the $Q\bar Q$ dynamics can be described
by a two-body positronium-like Schr\"odinger equation 
equation of the form ($E\equiv \sqrt{s}-2 M_Q$)
\begin{eqnarray}
& &
\bigg(\,
-\frac{\vec\nabla^2}{M_Q} 
- \frac{\vec\nabla^4}{4\,M_Q^3} 
+ \bigg[\,
  V_{c}(\vec r)
  + V_{\mbox{\tiny BF}}(\vec r) + V_{\mbox{\tiny NA}}(\vec r)
\,\bigg]  
- E
\,\bigg)\,G(\vec r,\vec r^\prime, E) \, =
\nonumber\\[2mm] &  &
\mbox{\hspace{2cm}} =
 \delta^{(3)}(\vec r-\vec r^\prime) 
\,,
\label{Schroedingerfull}
\end{eqnarray}
containing the heavy
quark kinetic energy up to NNLO and the instantaneous
potentials~(\ref{Coulombpotential})--(\ref{NApotential}). In
Eq.~(\ref{Schroedingerfull}) $M_Q$ is defined as the pole mass.
If $\Lambda_{\tiny\rm QCD}$ were of order or even larger
than $M_Q v^2$, on the other hand, the coupling of the radiation gluon
with the heavy quark would become of order one and retardation effects
would
be NNLO. In this case a perturbative calculation of $\sigma^{thr}$
would be impossible at NNLO because retardation effects would be
non-perturbative. For this reason we have to make sure that the
condition $M_Q v^2 > \Lambda_{\tiny\rm QCD}$ is satisfied, if
the NNLO expression for $\sigma^{thr}$ derived in the following
section shall be trusted.
\section{Calculation of the Total Cross Section}
\label{sectioncalculation}
For simplicity we only consider the photon mediated total cross
section. The inclusion of the $Z$ exchange is trivial for the vector
current contributions. The contributions from the axial-vector current
can be easily implemented at NNLO because the axial-vector current
produces the $Q\bar Q$ pair in a
P-wave state which leads to a suppression $\propto v^2$ relative to
the vector current contribution. We start from
the fully covariant expression for the normalized total cross section 
($R_{Q\bar Q}\equiv\sigma^{thr}/\sigma(e^+e^-\to\mu^+\mu^-)$)
\begin{eqnarray}
R_{Q\bar Q}(q^2) & = &
\frac{4\,\pi\,Q_b^2}{q^2}\,\mbox{Im}\,[\,
-i\,\int\,dx\,e^{i\,q.x}\,
  \langle\, 0\,| T\,j^b_\mu(x) \, j^{b\,\mu}(0)\, |\, 0\,\rangle]
\nonumber\\[2mm] & \equiv &
\frac{4\,\pi\,Q_b^2}{q^2}\,\mbox{Im}\,[\,-i\,
\langle\, 0\,| T\, \tilde j^b_\mu(q) \,
 \tilde j^{b\,\mu}(-q)\, |\, 0\,\rangle]
\,,
\label{crosssectioncovariant}
\end{eqnarray}
and expand the electromagnetic current which produces/annihilates
the $Q\bar Q$ pair with c.m. energy $\sqrt{q^2}$ in terms of
${}^3\!S_1$ NRQCD currents up to dimension eight ($i=1,2,3$)
\begin{eqnarray}
\tilde j_i(q) & = & b_1\,\Big({\tilde \psi}^\dagger \sigma_i 
\tilde \chi\Big)(q) -
\frac{b_2}{6 M_Q^2}\,\Big({\tilde \psi}^\dagger \sigma_i
(\mbox{$-\frac{i}{2}$} 
\stackrel{\leftrightarrow}{\mbox{\boldmath $D$}})^2
 \tilde \chi\Big)(q) + \ldots
\,,
\nonumber\\[2mm]
\tilde j_i(-q) & = & b_1\,\Big({\tilde \chi}^\dagger \sigma_i 
\tilde \psi\Big)(-q) -
\frac{b_2}{6 M_Q^2}\,\Big({\tilde \chi}^\dagger \sigma_i
(\mbox{$-\frac{i}{2}$} 
\stackrel{\leftrightarrow}{\mbox{\boldmath $D$}})^2
 \tilde \psi\Big)(-q) + \ldots 
\,,
\label{currentexpansion}
\end{eqnarray}
where the constants $b_1$ and $b_2$ are short-distance coefficients
normalized to one at the Born level. Only the spatial components of
the currents contribute at the NNLO level.
Inserting expansion~(\ref{currentexpansion}) back into 
Eq.~(\ref{crosssectioncovariant}) leads to the nonrelativistic
expansion of the NNLO cross section
\begin{eqnarray}
R_{\mbox{\tiny NNLO}}^{\mbox{\tiny thr}}(E) & = &
\frac{\pi\,Q_Q^2}{M_Q^2}\,C_1(\mu_{\rm hard},\mu_{\rm fac})\,
\mbox{Im}\Big[\,
{\cal{A}}_1(E,\mu_{\rm soft},\mu_{\rm fac})
\,\Big]
\nonumber\\[2mm]
& & - \,\frac{4 \, \pi\,Q_Q^2}{3 M_Q^4}\,
C_2(\mu_{\rm hard},\mu_{\rm fac})\,
\mbox{Im}\Big[\,
{\cal{A}}_2(E,\mu_{\rm soft},\mu_{\rm fac})
\,\Big]
+ \ldots
\,,
\label{crosssectionexpansion}
\end{eqnarray}
where
\begin{eqnarray}
{\cal{A}}_1 & = & i\,\langle \, 0 \, | 
\, ({\tilde\psi}^\dagger \vec\sigma \, \tilde \chi)\,
\, ({\tilde\chi}^\dagger \vec\sigma \, \tilde \psi)\,
| \, 0 \, \rangle
\,,
\label{A1def}
\\[2mm]
{\cal{A}}_2 & = & \mbox{$\frac{1}{2}$}\,i\,\langle \, 0 \, | 
\, ({\tilde\psi}^\dagger \vec\sigma \, \tilde \chi)\,
\, ({\tilde\chi}^\dagger \vec\sigma \, (\mbox{$-\frac{i}{2}$} 
\stackrel{\leftrightarrow}{\mbox{\boldmath $D$}})^2 \tilde \psi)\,
+ \mbox{h.c.}\,
| \, 0 \, \rangle
\,.
\label{A2def}
\end{eqnarray}
The cross section is expanded in terms of a sum of absorptive parts of
nonrelativistic current correlators, each of them multiplied by a
short-distance coefficient. In fact, the right-hand side (RHS) of
Eq.~(\ref{crosssectionexpansion})
just represents an application of the factorization formalism proposed
in~\cite{Bodwin1}. The second term on the RHS of
Eq.~(\ref{crosssectionexpansion}) is suppressed by $v^2$, i.e.\ of
NNLO. This can be seen explicitly by using the equations of motion
from the NRQCD Lagrangian, which relates the correlator
${\cal{A}}_2$ directly to ${\cal{A}}_1$,
\begin{equation}
{\cal{A}}_2 = M_Q\,E\,{\cal{A}}_1
\,.
\label{A2toA1}
\end{equation}
Relation~(\ref{A2toA1}) has also been used to obtain the coefficient
$-4/3$ in front of the second term on the RHS of
Eq.~(\ref{crosssectionexpansion}).
The nonrelativistic current correlators ${\cal{A}}_{1,2}$ 
contain the resummation of the singular terms mentioned
previously. They depend on the renormalization scale 
$\mu_{\rm soft}$ through the
potentials~(\ref{Coulombpotential})--(\ref{NApotential}).
The constants $C_1$ and $C_2$ (which are also normalized to one at
the Born level), on the other hand, describe short-distance effects
and, therefore, depend on the hard scale $\mu_{\rm hard}$. They only
represent a simple power series in $\alpha_s$ (where the coefficients
contain numbers and logarithms of $M_Q$, $\mu_{\rm fac}$ and
$\mu_{\rm hard}$) and do not contain any
resummations in $\alpha_s$. At NNLO they have to be calculated up to
order $\alpha_s^2$ because we count $\alpha_s/v$ of order one for a
perturbative nonrelativistic $Q\bar Q$ system.

The nonrelativistic correlators ${\cal{A}}_{1,2}$ are
calculated by determining the Green function of the Schr\"odinger
equation~(\ref{Schroedingerfull})
where $V_{\mbox{\tiny BF}}$ is evaluated for the ${}^3\!S_1$ configuration.
The NNLO relation between the correlator ${\cal{A}}_1$ and Green
function reads
\begin{eqnarray}
{\cal{A}}_1 & = & 6\,N_c\,\Big[\,
\lim_{|\vec r|,|\vec r^\prime|\to 0}\,G(\vec r,\vec r^\prime, E)
\,\Big]
\,.
\label{A1Greenfunctionrelation}
\end{eqnarray}
Eq.~(\ref{A1Greenfunctionrelation}) can be quickly derived from the
facts that $G(\vec r,\vec r^\prime,\tilde E)$ describes the propagation
of a quark-antiquark pair which is produced and annihilated at
relative distances $|\vec r|$ and $|\vec r^\prime|$, respectively, and
that the $Q\bar Q$ pair is produced and annihilated
through the electromagnetic current at zero
distances. Therefore ${\cal{A}}_1$ must be proportional to $\lim_{|\vec
r|,|\vec r^\prime|\to 0}\,G(\vec r,\vec r^\prime, E)$. The correct
proportionality constant can then be determined by considering
production of a free (i.e.\ $\alpha_s=0$) $Q\bar Q$ pair in the
nonrelativistic limit. (In this case the Born cross section in full QCD
can be easily compared to the imaginary part of
the Green function of the free nonrelativistic Schr\"odinger equation.)
The correlator ${\cal{A}}_2$ is determined from ${\cal{A}}_1$ via
relation~(\ref{A2toA1}).
We would like to emphasize that the zero-distance Green function on
the RHS of Eqs.~(\ref{A1Greenfunctionrelation}) contains UV
divergences from the higher dimensional NNLO effects which have to be
regularized. In the actual calculations carried out in 
Refs.~\cite{Hoang1,Hoang2}
we have imposed the explicit short-distance cutoff $\mu_{\rm fac}$,
called factorization scale. This is the reason why the correlators
also depend on $\mu_{\rm fac}$. 
One way to solve Eq.~(\ref{Schroedingerfull}) is to start from the well
known Green function $G_c^{(0)}$ of the nonrelativistic Coulomb
problem and to incorporate all the higher order terms via first and
second order Rayleigh-Schr\"odinger time-independent perturbation
theory. 

To determine the short-distance constant $C_1$ up to
${\cal{O}}(\alpha_s^2)$ we can expand
expression~(\ref{crosssectionexpansion})
in the (formal) limit $\alpha_s\ll v\ll 1$ (for
$\mu_{\rm soft}=\mu_{\rm hard}$) up to
${\cal{O}}(\alpha_s^2)$ and demand equality  (i.e.\ match) 
to the total cross section obtained at the two-loop level in full QCD
keeping terms up to NNLO in an expansion in $v$. In this limit fixed
multi-loop perturbation theory (i.e.\ an expansion in $\alpha_s$) as
well as the nonrelativistic approximation (i.e.\ a subsequent expansion
in $v$) are feasible. 
Because the full QCD cross section is independent of $\mu_{\rm fac}$,
$C_1$ also depends on $\mu_{\rm fac}$.
We call this kind
of matching calculation at the level of the final result
``direct maching''. The consistency of the
effective field theory approach ensures that $C_1$ only contains
contributions from momenta of order $M_Q$ and does not have any terms
singular in $v$. Details of this calculation and references
regarding the important calculations of the two-loop cross section in
full QCD at NNLO in the velocity expansion can be found in
Ref.~\cite{Hoang1}. A very economical method to calculate Feynman
diagrams in full QCD in an expansion in $v$ using dimensional
regularization, the ``Threshold Expansion'', has been developed in
Ref.~\cite{Smirnov1}.
\section{Applications}
\label{sectionapplications}
\subsection{Bottom Quark Mass from $\Upsilon$ Mesons}
Due to causality, 
derivatives of the electromagnetic current-current correlator with
respect to $q^2$ at $q^2=0$ are directly related to the total photon
mediated cross section of bottom quark-antiquark production in
$e^+e^-$ annihilation,
\begin{equation}
P_n \, \equiv \,
\frac{4\,\pi^2\,Q_b^2}{n!\,q^2}\,
\bigg(\frac{d}{d q^2}\bigg)^n\,\Pi_\mu^{\,\,\,\mu}(q)\bigg|_{q^2=0}
\, = \,
\int \frac{d s}{s^{n+1}}\,R_{b\bar b}(s)
\,.
\label{momentdef}
\end{equation}
\\[-2mm]
\noindent
\begin{figure}[t]
\begin{center}
\leavevmode
\epsfxsize2.5cm
\epsffile[220 420 420 550]{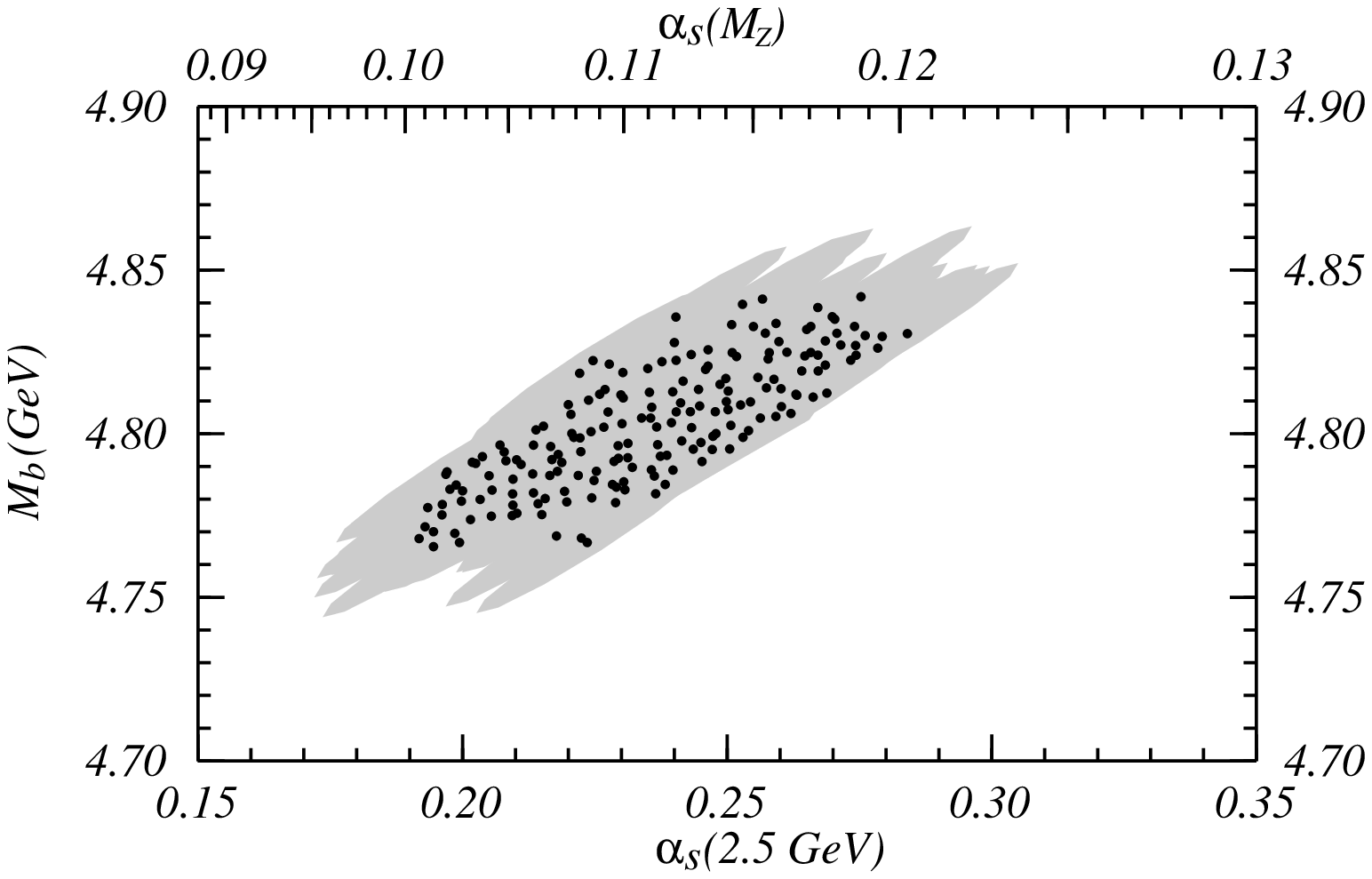}
\\
\vskip 2cm
\leavevmode
\epsfxsize2.5cm
\epsffile[220 420 420 550]{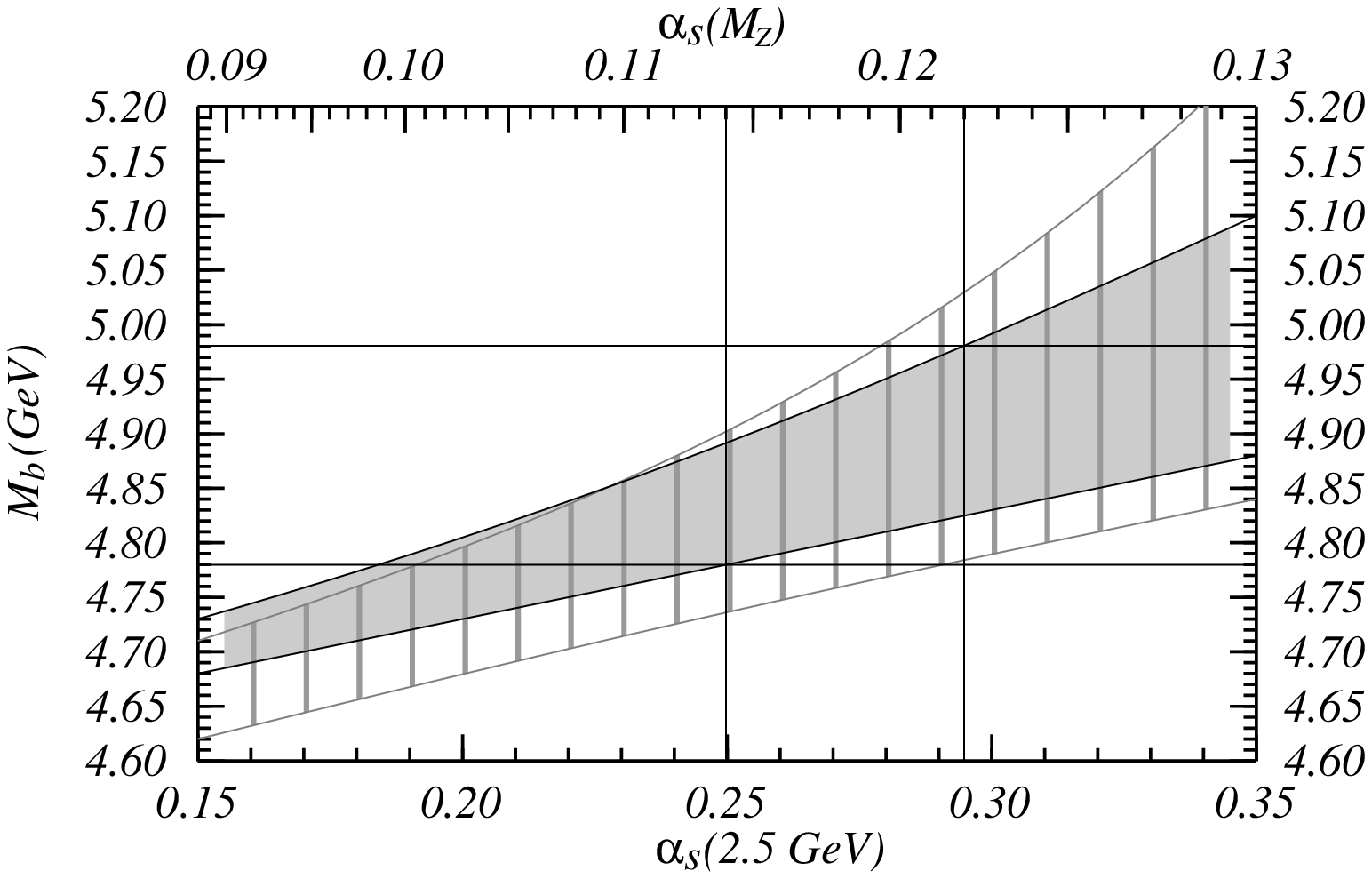}
\vskip  2.2cm
\end{center}
\caption{\label{fignnlofull} 
(a) Allowed region in the $M_b^{\tiny\rm pole}$-$\alpha_s$
plane for the unconstrained fit using the moments at NNLO. The gray
shaded region represents the allowed region. The dots are points
of minimal $\chi^2$ for a large number of models. Experimental errors
are included at the $95\%$ CL level and have been drawn as narrow
gray ellipses centered at the dots of minimal $\chi^2$.
(b) Allowed $M_b^{\tiny\rm pole}$ values for a given value
of $\alpha_s$.
The gray shaded region corresponds to the allowed ranges for the NNLO
analysis and the striped region for the NLO analysis. Experimental
errors are included at the $95\%$ CL level. It is illustrated how the
allowed range for $M_b^{\tiny\rm pole}$ at NNLO is obtained if
$0.114\le\alpha_s(M_z)\le 0.122$ is taken as an input. 
}
\end{figure}
Assuming global duality,
$P_n$ can be either calculated from experimental data for the total
cross section in $e^+e^-$ annihilation\footnote{
At the level of precision in this work the $Z$ mediated cross section
can be safely neglected.
}
or theoretically using quantum chromodynamics (QCD). It is the basic
idea of this sum rule to set the moments calculated from experimental
data, $P_n^{ex}$, equal to those determined theoretically from QCD,
$P_n^{th}$, and to use this relation to determine the bottom quark
mass (and the strong coupling) by fitting theoretical and experimental
moments for various values of $n$.
At this point it is important to set the range of allowed values of
$n$ for which the moments $P_n^{th}$ can be trusted using the NNLO cross
section calculated in the preceding section. As mentioned, we have to
make sure that $M_b \langle v\rangle^2 > \Lambda_{\tiny\rm QCD}$ where
$\langle v\rangle$ is the effective c.m. bottom quark velocity in a
particular moment. One can
show that $\langle v\rangle \sim 1/\sqrt{n}$  for large $n$. This means
that $n$ should be sufficiently smaller than $15-20$. It is 
interesting that the same conclusion can be drawn from the
Poggio-Quinn-Weinberg argument that the effective energy smearing range
contained in the moments should be larger than $\Lambda_{\tiny\rm
QCD}$~\cite{Poggio1}. On the other
hand, $n$ has to be large enough that the use of the cross section at
threshold is justified because the energy regime
close to threshold dominates $P_n$ only in this case. In our
analysis~\cite{Hoang1} we
have taken the range $4\le n\le 10$. Larger values of $n$ increase
the danger of large systematic errors. 
\begin{table}[t]
\caption{Recent determinations of bottom quark masses
using QCD sum rules for the $\Upsilon$ mesons. $m_b(m_b)$ refers to
the $\overline{\mbox{MS}}$ mass.
NLO refers to analyses including corrections of order $\alpha_s$
to the nonrelativistic limit and NNLO to analyses including
corrections or order $\alpha_s^2$, $\alpha_s v$ and $v^2$. No order is
indicated for Ref.~10  because the bound state poles have not
been taken into account for $P_n^{th}$ in that analysis.
For entries where also values of $\alpha_s$ are given a simultaneous
fit for mass and QCD coupling has been carried out.
In our analysis (Ref.~6) the errors are not written as Gaussian errors
but as allowed ranges.
\label{tab1}}
\vspace{0.2cm}
\begin{center}
\footnotesize
\begin{tabular}{|c|l|c|c|c|}
\hline
authors & order & $n$ & $M_b^{\tiny\rm pole}[\mbox{GeV}]$ & 
  $m_b(m_b)[\mbox{GeV}]$ \\
\hline
\hline
Ref.~\cite{Hoang1} 
& NLO & $4-10$ & 
\begin{minipage}{3cm}
\begin{center}
  $4.64-4.92$ \\
  $\alpha_s(M_z)=.086-.132$ 
\end{center}
\end{minipage} 
 & \\
\hline
 & NNLO & $4-10$ & 
\begin{minipage}{3cm}
\begin{center}
  $4.74-4.87$ \\
  $\alpha_s(M_z)=.096-.124$
\end{center}
\end{minipage} 
 & $4.17-4.35$ \\
\hline
 & NNLO & $4-10$ & $4.78-4.98$
 & $4.16-4.33$ \\
\hline
\hline
Ref.~\cite{Voloshin1} 
& NLO & $8-20$ & 
\begin{minipage}{3cm}
\begin{center}
  $4.827\pm .007$ \\
  $\alpha_s(M_z)=.109\pm .001$ 
\end{center}
\end{minipage} 
 & \\
\hline
Ref.~\cite{Jamin1} 
&   & $8-20$ & 
\begin{minipage}{3cm}
\begin{center}
  $4.604\pm .014$ \\
  $\alpha_s(M_z)=.118^{+.007}_{-.008}$ 
\end{center}
\end{minipage} 
 & 
\begin{minipage}{3cm}
\begin{center}
  $4.13\pm .06$ \\
  $\alpha_s(M_z)=.120^{+.010}_{-.008}$
\end{center}
\end{minipage} 
\\
\hline
Ref.~\cite{Kuehn1} 
& NLO & $10-20$ & $4.75\pm .04$ & \\
\hline
Ref.~\cite{Penin1} 
& NNLO & $10-20$ & $4.78\pm .04$ & \\
\hline
Ref.~\cite{Melnikov1} 
& NNLO & $14-18$ & &
  $4.20\pm .10$ \\
\hline
Ref.~\cite{Penin2} 
& NNLO & $8-12$ & $4.80\pm.06$ & \\
\hline
\end{tabular}
\end{center}
\end{table}
To determine the allowed range for the bottom quark mass we have
fitted $P_n^{th}$ to $P_n^{ex}$ for various sets of $n$'s. It turns
out that the theoretical errors are much larger than the experimental
ones. The dominant theoretical errors come from the dependence of 
$P_n^{th}$ on the scale $\mu_{\rm soft}$. We have combined
experimental and theoretical errors by using the ``scanning''
method~\cite{Buras1} in which a large number of statistical fits is
carried out for various ``reasonable'' choices for the scales, each
called ``a model''. It is believed that this method represents a
conservative way to combine experimental and large theoretical errors.
Our result for a simultaneous fit for the bottom
pole mass and $\alpha_s$ is displayed in Fig.~\ref{fignnlofull}a. 
In Fig.~\ref{fignnlofull}b the result for the pole mass is displayed
if $\alpha_s$ is taken as an input. It is conspicuous that the
extracted values for the pole mass are quite different for both
methods. This variation could be explained from the fact that the pole
mass is not defined beyond an accuracy of $\Lambda_{\tiny\rm QCD}$ due to
its strong infrared
sensitivity.~\cite{massrenormalon,Hoang3}. It therefore
seems to
be more advantageous to extract a short-distance mass like the 
$\overline{\mbox{MS}}$ mass. Taking into account the strong
correlations between pole mass and $\alpha_s$ and using the two-loop
conversion formula between pole and $\overline{\mbox{MS}}$ mass we
find very good agreement in the mass determination for both methods
(see Tab.~\ref{tab1}). Using the moments at NLO
we also found errors which were much more conservative than the ones
given in the NLO analysis of Ref.~\cite{Voloshin1}.
In Tab.~\ref{tab1} we give a compilation of all recent sum rule
determinations of bottom quark masses based on experimental data from
the $\Upsilon$ mesons.
\subsection{Top Quark Pair Production Cross Section at Threshold}
\label{subsectiontop}
The measurement of the $t\bar t$ production lineshape at threshold
($\sqrt{s}\approx 2 M_t$) is among the first tasks of the Next Linear
Collider (NLC). Due to the large top quark width ($\Gamma_t(t\to W b)\approx
(G_F/\sqrt{2}) M_t^3/8 \pi\approx1.5$~GeV) which serves as a natural
infrared cutoff and smearing
mechanism one can in fact calculate the lineshape locally without
imposing any additional smearing. To be more specific, the effective
c.m. velocity $v_{\tiny\rm eff}$ of the top quarks is of order
$\sqrt{E^2+\Gamma_t^2}/M_t$ which means that $M_t v_{\tiny\rm eff}^2$
is larger than $\Lambda_{\tiny\rm QCD}$ for any nonrelativistic
energy. Physically this cutoff mechanism arises because the top quarks
decay weakly before hadronization effects can set in. This also leads
to the phenomenon that there are no individual narrow toponium
resonances because the latter are so broad that they are almost
completely smeared out~\cite{Fadin1}.
In other words, there will be no toponium spectroscopy as we know it
from charmonia or bottomonia. However, the large decay width allows
for remarkably accurate measurements of top quark properties without
the diluting effects of hadronization. It is believed 
that measurements at the $t\bar t$ threshold at the NLC
will provide the most precise determination of the top quark mass. In
view of the potentially sizeable NNLO relativistic corrections it is
mandatory to calculate and control the NNLO effects.
Our calculation of the NNLO cross section has been designed for stable
quarks and can, strictly speaking, not be used as the NNLO cross section
for $t\bar t$.\footnote{
As already mentioned, we also have not yet included the Z exchange
contributions.
} 
Nevertheless we can illustrate the impact of the NNLO corrections by
using the naive replacement $E\to E+i \Gamma_t$ in
Eq.~(\ref{crosssectionexpansion}) where $\Gamma_t$ is the free top
quark width. This prescription has been proven to be correct in the
nonrelativistic limit~\cite{Fadin1} and should be sufficient to give
us the bulk of the NNLO relativistic corrections to the $t\bar t$
cross section at threshold.
\begin{figure}[t]
\begin{center}
\leavevmode
\epsfxsize3.5cm
\epsffile[220 420 420 550]{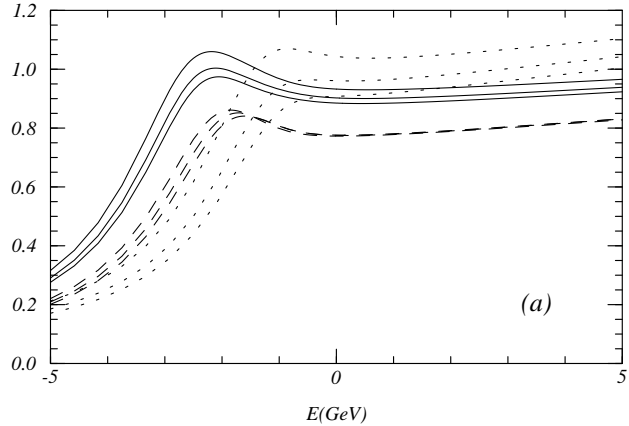}
\vskip  3.0cm
\end{center}
 \caption{\label{figtop} 
The total normalized photon-mediated $t\bar t$ cross section 
$R_{t\bar t}$ at
LO (dotted lines), NLO (dashed lines) and NNLO (solid lines)
in the nonrelativistic expansion
for the soft scales $\mu_{\rm soft}=50$ (upper), $75$ (middle)
and $100$~GeV (lower). The plot has been calculated in
Ref.~7. Similar results have been obtained in Ref.~20 and 21.}
\end{figure}
In Fig.~\ref{figtop} the LO (dotted lines), NLO (dashed lines) and NNLO
(solid lines) normalized cross sections are plotted versus
$E=\sqrt{q^2}-2M_t^{\tiny\rm pole}$ in the range 
$-5$~GeV$\,< E < 5$~GeV for $M_t^{\tiny\rm pole}=175$~GeV,
$\alpha_s(M_z)=0.118$ and
$\Gamma_t=1.43$~GeV. For the scales the choices 
$\mu_{\rm soft} = 50$ (upper lines), $75$ (middle lines) and $100$~GeV
(lower lines) and
$\mu_{\rm hard} = \mu_{\rm fac} = M_t$ have been
made and two-loop running of the strong coupling has been used.
It is evident that the NNLO corrections are large. 
The behavior of the NNLO corrections
clearly indicates that the convergence of the perturbative series for
the $t\bar t$ cross section is much worse than expected from the
general arguments given by Fadin and Khoze~\cite{Fadin1}.
Further examinations of the NNLO cross section including a study of
the impact of the use of a mass other than the pole one and a proper
treatment of the top width are mandatory.  
\section{Conclusions}
\label{sectionconclusions}
During the last year there has been significant progress in the
understanding of perturbative heavy quark-antiquark systems in the
kinematic regime close to threshold. Using the
concepts of effective field theories our knowledge 
has increased at the conceptual level and a number of previously
unknown NNLO corrections have been determined. In this talk I have
reviewed the ideas involved to
perturbatively calculate and then apply
the NNLO corrections of the total $Q\bar Q$ production cross section
at threshold. 
\section*{Acknowledgments}
I am very grateful to P.~Labelle, T.~Teubner and S.~M.~Zebarjad for
their collaboration on topics reported here. 
I would like to thank the Theoretical Physics Institute for financial 
support and the warm hospitality during the conference and, in
particular M.~B.~Voloshin and A.~Vainshtein for fruitful discussions.
I thank Z.~Ligeti for reading the manuscript. This work is supported
in part by the U.S.~Department of Energy under contract
No.~DOE-FG03-97ER40546.



\section*{References}
\begin{thebibliography}{99}

\bibitem{Caswell1} W. E. Caswell and G. E. Lepage, 
     \Journal{\PLB}{167}{437}{1986}. 

\bibitem{Pineda1} A. Pineda and J. Soto, 
     {\em Nucl. Phys. Proc. Suppl. }{\bf 64}, 428 (1998).

\bibitem{Fischler1} W. Fischler, \Journal{\NPB}{129}{157}{1977};
     A. Billoire, \Journal{\PLB}{92}{343}{1980}.

\bibitem{Peter1} M. Peter, \Journal{\PRL}{78}{602}{1997}; 
     \Journal{\NPB}{501}{471}{1997}.

\bibitem{Bodwin1} G. T. Bodwin, E. Braaten, and G. P. Lepage, 
     \Journal{\PRD}{51}{1125}{1995}.

\bibitem{Hoang1} A. H. Hoang, UCSD/PTH 98-02 [hep-ph/9803454].

\bibitem{Hoang2} A. H. Hoang and T. Teubner, UCSD/PTH 98-01
     [hep-ph/9801397].

\bibitem{Smirnov1} M. Beneke and V.A. Smirnov, 
     \Journal{\NPB}{522}{321}{1998}.

\bibitem{Voloshin1} M. B. Voloshin, 
     {\em Int. J. Mod. Phys. }{\bf A 10}, 2865 (1995).

\bibitem{Jamin1} M. Jamin and A. Pich, 
     \Journal{\NPB}{507}{334}{1997}.

\bibitem{Kuehn1} J. H. K\"uhn, A. A. Penin, and A. A. Pivovarov,
     TTP/98-01 [hep-ph/9801356].

\bibitem{Penin1} A. A. Penin and A. A. Pivovarov, TTP/98-13
     [hep-ph/9803363].

\bibitem{Melnikov1} K. Melnikov and A. Yelkovsky, TTP/98-17 
     [hep-ph/9805270].

\bibitem{Penin2} A. A. Penin and A. A. Pivovarov, 
     INR-98-0986 [hep-ph/9807421].

\bibitem{Poggio1} E. C. Poggio, H. R. Quinn, and S. Weinberg, 
     \Journal{\PRD}{13}{1958}{1976}.

\bibitem{Buras1} A. J. Buras, proceedings of the workshop {\em
     Symposium of Heavy Flavours}, Santa Barbara, July 7--11, 1997
     [hep-ph/9711217].

\bibitem{massrenormalon}  M. Beneke and V. M. Braun, 
     \Journal{\NPB}{426}{301}{1994};\\
     I. I. Bigi {\it et al.}, \Journal{\PRD}{50}{2234}{1994}.

\bibitem{Hoang3} A. H. Hoang, M. C. Smith, T. Stelzer and
     S. Willenbrock, UCSD/PTH 98-13 [hep-ph/9804227];\\
     M. Beneke, CERN-TH/98-120 [hep-ph/9804241].

\bibitem{Fadin1} V. S. Fadin and V. A. Khoze, 
     {\em Sov. J. Nucl. Phys. }{\bf 48}, 309 (1988).

\bibitem{Melnikov2} K. Melnikov and A. Yelkovsky, BudkerINP-98-7
     [hep-ph/9802379].

\bibitem{Yakovlev1} O. Yakovlev, WUE-ITP-98-036
     [hep-ph/9808463].

\end{thebibliography}
\end{document}